\documentclass[11pt,headsepline]{scrartcl}

\usepackage[latin1]{inputenc}   
\usepackage{latexsym}
\usepackage{amssymb}
\usepackage{amsfonts}
\usepackage{amsmath}
\usepackage[numbers,sort&compress]{natbib}
\usepackage{bm}
\usepackage{graphicx}
\usepackage{hyperref}
\usepackage{color}
\usepackage{multicol}

 \providecommand{\dprod}{\!\cdot\!}%
 \providecommand{\wprod}{\!\wedge \!}
 
 \providecommand{\sfrac}[2]{\frac{#1}{#2}\,}
 \providecommand{\email}[1]{email: \href{mailto:#1}{\texttt{#1}}}

\numberwithin{equation}{section}

\begin{document}

\newcommand{\mytitle}{Different algebras for one reality}
\title{\mytitle}

\author{José B. Almeida\\ Universidade do Minho, Physics
Department\\
Braga, Portugal, \email{bda@fisica.uminho.pt}}

\date{}

\pagestyle{myheadings} \markright{José B. Almeida \hfill \mytitle}


\maketitle

\begin{abstract}
The most familiar formalism for the description of geometry
applicable to physics comprises operations among 4-component vectors
and complex real numbers; few people realize that this formalism has
indeed 32 degrees of freedom and can thus be called 32-dimensional.
We will revise this formalism and we will briefly show that it is
best accommodated in the Clifford or geometric algebra
$\mathcal{G}_{1,3}\times \mathbb{C}$, the algebra of 4-dimensional
spacetime over the complex field.

We will then explore other algebras isomorphic to that one, namely
$\mathcal{G}_{2,3}$, $\mathcal{G}_{4,1}$ and $\mathbb{Q}\times
\mathbb{Q}\times\mathbb{C}$, all of which have been used in the past
by PIRT participants to formulate their respective approaches to
physics. $\mathcal{G}_{2,3}$ is the algebra of 3-space with two time
dimensions, which John Carroll used implicitely in his formulation
of electromagnetism in $3 + 3$ spacetime, $\mathcal{G}_{4,1}$ was
and it still is used by myself in a tentative to unify the
formulation of physics and $\mathbb{Q}\times
\mathbb{Q}\times\mathbb{C}$ is the choice of Peter Rowlands for his
nilpotent formulation of quantum mechanics. We will show how the
equations can be converted among isomorphic algebras and we also
examine how the monogenic functions that I use are equivalent in
many ways to Peter Rowlands nilpotent entities.

\vspace{12pt} \noindent PACS numbers: {04.50.-h; 02.40.-k.}
\end{abstract}


\section{Introduction}
We call Physics to a discipline that creates mathematical models of
physical reality. In practice, we write mathematical equations whose
solutions allow us to predict the outcome of experiments and
observations. One physical model is just as good as the predictions
it allows and the most successful models become known as physical
theories. Every model makes use of a limited set of independent
variables, which can be operated among themselves; we say that the
model uses an underlying algebra. The model must also give physical
meaning to the independent variables and algebraic operations
performed among them, so that everybody can then translate into
reality the results of operations performed within the model.

In view of what was said above, one sees that an algebra is an
intrinsical component of any physical model, but it happens quite
often that several algebras are only apparently different and can be
shown to be isomorphic to each other. When this happens, models
incorporating such algebras are frequently equivalent, although the
insight one has over problems addressed with two equivalent models
may be entirely different. In the following sections we will discuss
the algebras associated with models proposed by various authors,
showing that they are in many cases isomorphic. We will also show
how to convert equations between isomorphic algebras. In the case of
a model proposed by John Carroll \cite{Carroll04}, considering a
space with 3 spatial and 3 temporal dimensions, the associated
algebra is a superalgebra of several 5-dimensional algebras, so, the
isomorphisms that can be found apply only to a subalgebra of the one
proposed by the author.

\section{Algebras most frequently used in physics}
Both Newtonian mechanics and Maxwell's equations are models based on
4 independent variables, 3 space coordinates and 1 scalar time
variable. The algebras used to operate with these variables are the
algebras of real and complex numbers complemented with vector
algebra, but it is easy to see that this system lacks consistency.
For instance, two vectors $\mathbf{a}$ and $\mathbf{b}$ determine a
parallelogram with area given by $|\mathbf{a} \times \mathbf{b}| =
|\mathbf{b} \times \mathbf{a}|$. We make use of an operation among
two vectors and then define the area as a scalar quantity. It makes
more sense to define a new product whose outcome is an oriented
area, called \emph{outer product} and denoted $\mathbf{a} \wprod
\mathbf{b}$. The outcome of the outer product is precisely the area
of the parallelogram defined by the two vectors, with a sign defined
by the direction of movement from one vector to the other.

Clifford algebras are based on the \emph{geometric product} or
simply the product of vectors, incorporating both the inner and
outer products. For any two vectors it is
\begin{equation}
    \mathbf{a}\mathbf{b} = \mathbf{a} \dprod \mathbf{b} + \mathbf{a}
    \wprod \mathbf{b}.
\end{equation}
The geometric product is associative and so it is possible to have
products of 3 vectors, leading to a grade-$3$ element of the type
$\mathbf{a} \wprod \mathbf{b} \wprod \mathbf{c}$ which, if not zero,
represents an oriented volume. We can thus say that the algebra
associated with the spatial part of Newtonian mechanics and
Maxwell's equations is Clifford algebra of dimension 3, also known
as \emph{geometric algebra} of dimension 3 and denoted
$\mathcal{G}_{3}$ or $\mathcal{G}_{3,0}$. The volume elements of
this algebra, as well as the highest grade elements of any Clifford
algebra, are called \emph{pseudoscalars}. For an extensive treatment
of geometric algebras see \cite{Doran03, Hestenes84}.

Further problems with the Newtonian and Maxwellian models reside in
the fact that time is treated as scalar but it has to be
differentiated from the scalar coefficients of vectors. This is
solved in special relativity, because it proposes that time is to be
treated as a dimension of spacetime, thus increasing the
dimensionality of the associated geometric algebra to 4; the highest
grade element is now a 4-dimensional hypervolume. There are two
possible algebras, $\mathcal{G}_{1,3}$ and $\mathcal{G}_{3,1}$,
associated with one positive and 3 negative norm frame vectors or
one negative and 3 positive norm frame vectors, respectively. The
former is the most common choice and it allows the formulation of
most physics equations, including quantum mechanics
\cite{Hestenes03:2}. In order to fully accommodate quantum mechanics
one must, however, allow for complex coefficients, a possibility not
considered in \cite{Hestenes03:2}.

Starting with the work by Theodor Kaluza, who proposed a
5-dimensional unification of electromagnetism with general
relativity \cite{Kaluza21}, some authors have used higher
dimensional spaces to try and unify the equations of physics. My own
work makes use of 5-dimensional spacetime and bears a strong
relation to Kaluza's \cite{Almeida04:4, Almeida06:4}. The geometric
algebra associated with 5-dimensional spacetime in this formulation
is $\mathcal{G}_{4,1}$ but other authors have used the opposite
signature $\mathcal{G}_{1,4}$ \cite{Sehara01}. How different and how
similar are all these approaches?

In order to answer the question we start by examining the overall
dimensionality of the different algebras, starting with the algebra
of physical space, $\mathcal{G}_{3,0}$. We realize that the elements
of the algebra can be classified into 4 grades: scalars, vectors,
areas and volumes, or better, grades $0$, $1$, $2$ and $3$. While
both scalars and volumes have no associated orientation besides
positive and negative, vectors and areas have 3 possible
orientations, so, the total number of degrees of freedom is 8 and we
say that total dimensionality is 8. In a similar way, the total
dimensionality of a general geometric algebra, $\mathcal{G}_{p,q}$
is $2^{p+q}$, if only real coefficients are allowed for all grades.
If complex coefficients are allowed the total dimensionality is
either doubled or remains unaltered relative to the real coefficient
version. Some algebras can be classified as \emph{complex algebras},
because their pseudoscalar elements have negative square and commute
with all other elements. In complex algebras the unit pseudoscalar
doubles as the complex imaginary, so, introducing complex
coefficients does not bring in any extra dimensions. In non-complex
algebras the introduction of complex coefficients doubles the
degrees of freedom, doubling the total dimensionality.

All geometric algebras are isomorphic to one particular matrix
algebra, over one particular field that provides the coefficients.
What this means is that all operations performed in a particular
geometric algebra have equivalent operations in the isomorphic
matrix algebra. The use of matrix algebra isomorphism is useful for
classification purposes, but it is usually not recommended for
performing operations since all the links with geometry are lost.
\begin{table}
\caption{\label{t:representation}Matrix representation of Clifford
Algebras $\mathcal{C}\ell(p,q)$, with $p$ positive and $q$ negative
norm frame vectors. The notation $\mathbb{F}(n)$ is used for the
$n$-dimensional matrix algebra over the field $\mathbb{F}$ and $\,^2
\mathbb{F}(n)$ identifies the sum $\mathbb{F}(n) \bigoplus
\mathbb{F}(n)$; $\mathbb{R}$ stands for real numbers, $\mathbb{C}$
for complex numbers and $\mathbb{Q}$ for quaternions.}
\begin{center}
\begin{tabular}{c|cccccccc}
   \begin{tabular}{lr}
   & q \\
   p &
   \end{tabular}
 &  0    &     1    &     2    &     3    &    4    &     5    & 6
&
  7\\
  \hline \\
  0   & $    \mathbb{R}   $ & $    \mathbb{C}   $ & $    \mathbb{Q}   $ & $   \,^2 \mathbb{Q}    $ & $   \mathbb{Q}(2)  $ & $  \mathbb{C}(4)  $ & $  \mathbb{R}(8) $ & $
  \,^2 \mathbb{R}(8) $ \\
1   & $   \,^2 \mathbb{R}   $ & $    \mathbb{R}(2)  $ & $
\mathbb{C}(2) $ & $ \mathbb{Q}(2) $ & $  \,^2 \mathbb{Q}(2) $ & $
\mathbb{Q}(4) $ & $ \mathbb{C}(8) $ & $
\mathbb{R}(16)  $ \\
2   & $ \mathbb{R}(2) $ & $  \,^2 \mathbb{R}(2) $ & $ \mathbb{R}(4)
$ & $ \mathbb{C}(4)  $ & $ \mathbb{Q}(4) $ & $ \,^2 \mathbb{Q}(4) $
& $ \mathbb{Q}(8) $ & $
\mathbb{C}(16) $ \\
3  & $ \mathbb{C}(2) $ & $ \mathbb{R}(4) $ & $  \,^2 \mathbb{R}(4) $
& $ \mathbb{R}(8) $ & $ \mathbb{C}(8)  $ & $ \mathbb{Q}(8) $ & $
\,^2 \mathbb{Q}(8) $ & $ \mathbb{Q}(16) $
\\
 4  & $ \mathbb{Q}(2) $ & $ \mathbb{C}(4) $ & $ \mathbb{R}(8) $ & $ \,^2 \mathbb{R}(8)  $ & $  \mathbb{R}(16) $ & $  \mathbb{C}(16) $ & $  \mathbb{Q}(16) $ & $ \,^2 \mathbb{Q}(16) $ \\
  5  & $ \,^2 \mathbb{Q}(2) $ & $ \mathbb{Q}(4) $ & $ \mathbb{C}(8) $ & $ \mathbb{R}(16) $ & $ \,^2 \mathbb{R}(16) $ & $ \mathbb{R}(32) $ & $  \mathbb{C}(32) $ & $ \mathbb{Q}(32) $  \\
  6  & $ \mathbb{Q}(4) $ & $  \,^2 \mathbb{Q}(4)  $ & $  \mathbb{Q}(8) $ & $ \mathbb{C}(16) $ & $ \mathbb{R}(32) $ & $ \,^2 \mathbb{R}(32) $ & $ \mathbb{R}(64) $ & $ \mathbb{C}(64) $ \\
  7  & $ \mathbb{C}(8) $ & $   \mathbb{Q}(8) $ & $  \,^2 \mathbb{Q}(8) $ & $   \mathbb{Q}(16) $ & $ \mathbb{C}(32) $ & $
\mathbb{R}(64) $ & $ \,^2 \mathbb{R}(64) $ & $ \mathbb{R}(128) $

\end{tabular}
\end{center}
\end{table}
Table \ref{t:representation} shows the matrix algebras isomorphic to
the lowest order geometric algebras. The entries in the table are of
the type $\mathbb{F}(n)$, which stands for algebra of
$n$-dimensional matrices with coefficients in the field
$\mathbb{F}$. The coefficients' field can be real numbers
($\mathbb{R}$), complex numbers ($\mathbb{C}$) or quaternions
($\mathbb{Q}$). A few algebras are non-simple and are denoted $\,^2
\mathbb{F}(n)$; this means that two copies of the $\mathbb{F}(n)$
algebra are needed in the isomorphism. Looking up the table for the
matrix representation of physical space algebra,
$\mathcal{G}_{3,0}$, we see that we must use 2-dimensional matrices
with complex coefficients. Usually we associate the frame vectors
$\{\sigma_m \}$ to the Pauli matrices, as follows:
\begin{equation}
    \sigma_1 \equiv \begin{pmatrix} 0 & 1 \\ 1 & 0 \end{pmatrix},
    \quad
    \sigma_2 \equiv \begin{pmatrix} 0 & -\mathrm{i}\\ \mathrm{i} & 0 \end{pmatrix},
    \quad
    \sigma_3 \equiv \begin{pmatrix} 1 & 0 \\ 0 & -1 \end{pmatrix}.
\end{equation}
Under the matrix isomorphism scalars are represented by the product
of a real number by the identity matrix, vectors by linear
combinations of matrices $\sigma_m$, areas by linear combinations of
two Pauli matrix products and volumes by the product of a real
number by $\hat{\sigma}_1 \hat{\sigma}_2 \hat{\sigma}_3$, the
notation $\hat{\sigma}_m$ being used for matrices. Since the product
of the three Pauli matrices is the identity matrix multiplied by the
complex imaginary, we see that the unit pseudoscalar of the algebra
actually doubles as imaginary.

Minkowski spacetime is most frequently associated with
$\mathcal{G}_{1,3}$ algebra, although several authors prefer the
$\mathcal{G}_{3,1}$ alternative. No physical significance is
attributed to the choice of signature, but one sees from Table
\ref{t:representation} that the corresponding algebras are not
isomorphic; there is probably some deep meaning in this choice that
has escaped physicists so far. For the matrix representation of
$\mathcal{G}_{3,1}$, the most direct route starts with Majorana
gamma matrices, which have only imaginary elements, proceeding to
assign the four frame vectors from the algebra by the equation
\begin{equation}
    \sigma_\mu \equiv \mathrm{i} \hat{\gamma}_\mu;
\end{equation}
the notation $\hat{\gamma}_\mu$ is used here for matrices. For the
$\mathcal{G}_{1,3}$ algebra we should, in principle, select Pauli
matrices $\hat{\sigma}_1$ and $\hat{\sigma}_3$ over the quaternion
field. There is a workaround that avoids the discomfort of
quaternions, which consists on allowing for 4-dimensional matrices
with complex elements and restricting the matrix coefficients to
real numbers. There several possible alternatives for the assignment
of basis vectors to matrices, the most common being derived from
Dirac-Pauli representation; this is
\begin{equation}
\label{eq:diracmatrix}
    \gamma_0 \equiv \begin{pmatrix}
    I & 0 \\ 0 & -I
    \end{pmatrix}, \quad
    \gamma_m \equiv \begin{pmatrix}
    0 & \hat{\sigma}_m \\ -\hat{\sigma}_m & 0
    \end{pmatrix}.
\end{equation}
These matrices have both real and imaginary elements, but used with
real coefficients they still provide a basis representation for
$\mathcal{G}_{1,3}$, avoiding the use of quaternions.

In 5-dimensional spacetime the representation is much easier with
$\mathcal{G}_{4,1}$ then with $\mathcal{G}_{1,4}$, because the
latter not only needs quaternions but it is also a non-simple
algebra; we will not pay much attention to this case. With
$\mathcal{G}_{4,1}$ we have a beautiful scenario; we can use
4-dimensional matrices with complex elements and complex
coefficients. Among the various possible assignments we propose the
following one, which is derived from the Dirac-Pauli representation,
as we shall see below:
\begin{equation}
\label{eq:G_4_1}
\begin{split}
    e_0 \equiv   \begin{pmatrix} 0 & 0 & 1 & 0  \\
    0 & 0 & 0 & 1 \\ -1 & 0 & 0 & 0 \\ 0 & -1 & 0 &
    0
    \end{pmatrix},~~
    e_1 \equiv & \begin{pmatrix} 0 & 1 & 0 & 0 \\
    1 & 0 & 0 & 0 \\ 0 & 0 & 0 & -1 \\ 0 & 0 & -1 & 0
    \end{pmatrix},~~
    e_2 \equiv   \begin{pmatrix} 0 & -\mathrm{i} & 0 & 0 \\
    \mathrm{i} & 0 & 0 & 0 \\ 0 & 0 & 0 & \mathrm{i} \\ 0 & 0 & -\mathrm{i} & 0
    \end{pmatrix} \\
    e_3 \equiv \begin{pmatrix} 1 & 0 & 0 & 0 \\
    0 & -1 & 0 & 0 \\ 0 & 0 & -1 & 0 \\ 0 & 0 & 0 & 1
    \end{pmatrix} & ,~~
    e_4 \equiv  \begin{pmatrix} 0 & 0 & -1 & 0 \\
    0 & 0 & 0 & -1 \\ -1 & 0 & 0 & 0 \\ 0 & -1 & 0 & 0
    \end{pmatrix} .
\end{split}
\end{equation}
We have not covered in this section the matrix representations for
Carroll's $\mathcal{G}_{3,3}$ \cite{Carroll04} or Rowlands'
$\mathbb{Q}\times \mathbb{Q}\times\mathbb{C}$ \cite{Rowlands03,
Rowlands07}, although the former can be looked up in the table. We
will consider these algebras in the next section.

\section{Converting equations among algebras}
We have seen in the previous section that there are isomorphisms
between the algebras of different spaces, which means that it is
feasible to translate all equations from one algebra to any of its
isomorphic algebras. Although the equations can be translated, the
geometric connection varies substantially an so does the insight one
has over the equations. As an example, take the Dirac equation,
which appears formulated as a matrix equation in every textbook. The
standard formulation does not allow any geometrical interpretation,
because matrices have no connection to geometry whatsoever. The fact
that Dirac equation can be translated into geometric algebra
provides the necessary link to geometry and the solutions can be
interpreted geometrically \cite{Almeida06:4}.

If all we are interested in is the formulation of general
relativity, 4-dimensional spacetime is the adequate choice, which
has a total dimensionality of 16. Physics equations, however,
involve the use of complex numbers, at least for quantum mechanics.
The total dimensionality implied by the set of physics equations for
general relativity and quantum mechanics is then 32 and our task is
then to translate equations among algebras with this total
dimensionality. We start with Dirac-Pauli matrices, as defined in
Eq.\ (\ref{eq:diracmatrix}), and we follow the usual procedure for
the definition of matrix $\hat{\gamma}_5$:
\begin{equation}
    \hat{\gamma}_5 = \mathrm{i} \hat{\gamma}_0 \hat{\gamma}_1
    \hat{\gamma}_2 \hat{\gamma}_3.
\end{equation}
The translation between Dirac algebra and the algebra of
5-dimensional spacetime, $\mathcal{G}_{4,1}$, is made directly by
the following relations
\begin{equation}
    e_\mu \equiv \hat{\gamma}_\mu \hat{\gamma}_5, \quad e_4 \equiv -\hat{\gamma}_5.
\end{equation}
This equation can be interpreted both as a matrix or a geometric
algebra equation. Indeed, if the $\gamma_\mu$ represent the frame
vectors of Minkowski spacetime, the equation can be read as a
geometric algebra equation and allows the transposition from
Minkowski into 5-dimensional spacetime. The inverse transposition
follows the rules:
\begin{equation}
\label{eq:dirac_5d}
    \hat{\gamma}_\mu \equiv e_4 e_\mu, \quad \hat{\gamma}_5 \equiv
    -e_4.
\end{equation}

We turn our attention now to Rowlands' algebra, whose elements are
sets of two quaternions and one complex number. For convenience we
shall represent a general element of this algebra with the notation
$\mathbf{q}\mathsf{q}c$; boldface and sanserif characters represent
two independent quaternions and a normal character represents a
complex number. The elements in the set can be commuted, so, the
total dimensionality of the algebra is $4 \times 4 \times 2 = 32$,
just as Dirac's algebra. The basis for Rowlands' algebra is given by
the sets
\begin{equation*}
\begin{split}
    & \{1, \mathbf{i}, \mathbf{j}, \mathbf{k} \},\\
    & \{1, \mathsf{i}, \mathsf{j}, \mathsf{k} \},\\
    & \{ \mathrm{i} \}.
\end{split}
\end{equation*}
The first quaternion basis verifies the relations
\begin{equation}
\begin{split}
    & \mathbf{i}^2 = \mathbf{j}^2 = \mathbf{k}^2 = -1,\\
    & \mathbf{i} \mathbf{j} = - \mathbf{j}\mathbf{i} = \mathbf{k};
\end{split}
\end{equation}
similar relations hold for the other quaternion. In order to set up
the conversion relations for $\mathcal{G}_{4,1}$ we start by
defining 3 anticommuting elements that can be associated with the 3
physical space dimensions; for this we set
\begin{equation}
    e_1 \equiv \mathbf{i} \mathsf{i}, \quad e_2 \equiv \mathbf{j} \mathsf{i},
    \quad e_1 \equiv \mathbf{k} \mathsf{i}.
\end{equation}
We note that the unit volume is now
\begin{equation}
    e_1 e_2 e_3 = \mathbf{i} \mathbf{j} \mathbf{k}\, \mathsf{i} = -\mathsf{i}
     .
\end{equation}
Now we need to find an element that anticommutes with the former
ones, with negative square, for $e_0$, and a second one, squaring to
unity, for $e_4$. A possible choice is
\begin{equation}
    e_0 = \mathsf{j}, \quad e_4 = \mathrm{i} \mathsf{k}.
\end{equation}
We need to check that the unit pseudoscalar coincides with the
complex imaginary, so, we do
\begin{equation}
    e_0 e_1 e_2 e_3 e_4 = \mathrm{i}\, \mathsf{j} \mathsf{i}\mathsf{k}
    = \mathrm{i}.
\end{equation}
The inverse relations are very easy to establish. With the help of
the above conversion relations it becomes a feasible task to convert
all equations between Dirac's, Rowlands' and my own notations but,
if physics is the same in all notations, the insight and
comprehension one has over the problems at hand can gain a lot from
different approaches.

The best equation to test the conversion relations is arguably the
Dirac equation; this is written, in terms of matrices, as
\begin{equation}
    \hat{\gamma}^\mu \partial_\mu \psi + \mathrm{i}m \psi = 0.
\end{equation}
Upper indices are used here and elsewhere to denote a change of
sign, with respect to the corresponding lower indices, for those
elements that square to $-1$ ($-I$ in the matrix case). Multiplying
on the left by $\hat{\gamma}^5$ and using conversion relations from
Eq.\ (\ref{eq:dirac_5d}), the Dirac equation becomes
\begin{equation}
    e^\mu \partial_\mu \psi + \mathrm{i}m \psi = 0.
\end{equation}
We now establish that $\mathrm{i}m \psi = \partial_4 \psi$, that is,
we establish that the wavefunction dependence on $x^4$ is harmonic
and is governed by the particle's mass. This is very similar to a
compactification of coordinate $x^4$. The Dirac equation acquires a
new form:
\begin{equation}
\label{eq:monogenic}
    e^\alpha \partial_\alpha \psi = \nabla \psi = 0.
\end{equation}
The index $\alpha$ runs from $0$ to $4$ and the symbol $\nabla$
represents what is known as the vector derivative of the algebra.
Any function $\psi$ that is a solution of Eq.\ (\ref{eq:monogenic})
is called monogenic. There are plane wave solutions for this
equation, with the general form
\begin{equation}
    \psi = \psi_0 \mathrm{e}^{\mathrm{i}(p_\alpha x^\alpha +
    \theta)}.
\end{equation}
The monogenic equation implies that $e^\alpha p_\alpha \psi_0 = 0$,
which can only be true if $(e^\alpha p_\alpha)^2 = 0$ and $\psi_0$
includes a factor $e^\alpha p_\alpha$. We say that the vector $p =
e^\alpha p_\alpha$ is a nilpotent vector. In the above cited works,
Rowlands uses the nilpotency condition as first principle, but we
see here how this can be derived from the monogenic condition. If
one establishes monogeneity as first principle, then the nilpotency
condition is implied.

In its matrix version, Dirac's equation accepts column matrix
solutions, which are called Dirac spinors. In order to find the
geometric equivalent of these we define 4 orthogonal idempotent
elements by the relations
\begin{equation}
\begin{split}
    f_1 &= \sfrac{1}{4} (1 + e_3)(1 + \mathrm{i} e_1 e_2), \\
    f_2 &= \sfrac{1}{4} (1 - e_3)(1 + \mathrm{i} e_1 e_2), \\
    f_3 &= \sfrac{1}{4} (1 - e_3)(1 - \mathrm{i} e_1 e_2), \\
    f_4 &= \sfrac{1}{4} (1 + e_3)(1 - \mathrm{i} e_1 e_2).
\end{split}
\end{equation}
These elements are called idempotents because their powers are
always equal to the element itself. They are orthogonal because the
product of any two different idempotents returns zero; They also add
to unity. We can then split the original monogenic function into
four components as in
\begin{equation}
    \psi = \sum_{i =1}^4 \psi f_i = \sum_{i =1}^4 \psi_i.
\end{equation}
Each of the terms $\psi_i$ is still a monogenic function and it is
the geometric version of a Dirac spinor. Rowlands' nilpotents have 4
components and they are also another form of spinors.

Now, the case of Carroll's $\mathcal{G}_{3,3}$ algebra
\cite{Carroll04} does not readily fall into the algebras we have
discussed above because, being 6-dimensional, it has a total
dimensionality of 64, doubling the dimensionality of those algebras.
However, Carroll argues that there is one special time dimension,
which corresponds to ordinary time, and two orthogonal time
dimensions, which must be treated differently. Carroll's proposed
wavefunction is the solution of the second order equation
\begin{equation}
    -(\partial_{s1}^2 + \partial_{s2}^2 + \partial_{s3}^2) \psi +
    m^2 \psi + \partial_{t3} \psi = 0;
\end{equation}
where
\begin{equation}
\label{eq:carroll}
    m^2 \psi = (\partial_{t1}^2 + \partial_{t2}^2) \psi.
\end{equation}
For the purpose of this equation we can define a combined time
coordinate, using $t1$ and $t2$, by
\begin{equation}
    tc = \sfrac{1}{2} (t1 + t2).
\end{equation}
Equation (\ref{eq:carroll}) is then a $\mathcal{G}_{2,3}$ algebra
equation and we see from Table \ref{t:representation} that this
algebra is isomorphic to $\mathcal{G}_{4,1}$. In order to convert
between the two algebras we define the vectors for
$\mathcal{G}_{2,3}$ by
\begin{equation}
\begin{split}
    e_{sm} &= \mathrm{i} e_m, \\
    e_{t3} &= \mathrm{i} e_0, \\
    e_{tc} &= e_4.
\end{split}
\end{equation}
With this conversion it is easy to verify that Eq.\
(\ref{eq:carroll}) is indeed a second order version of Eq.\
(\ref{eq:monogenic}). We don't discuss here other implications of
Carroll's 6-dimensional approach, the purpose of this discussion
being only to show that there is an implied 5-dimensional algebra
isomorphic to the other ones presented above.

\section{Conclusion}
Many authors resort to different algebras for the exposition of
their own approaches to fundamental physical equations, such as
Maxwell's equations, Dirac's equation and Einstein's equations.
Quite frequently authors propose their own versions of those
equations, highlighting the virtues of their approaches. The task of
comparing results is difficult because the form of both equations
and their solutions is dependent on the particular algebra that the
author has chosen. We have shown that the algebra used by Dirac has
an overall dimensionality of 32, the same as several 5-dimensional
algebras proposed by different authors. The tensor algebra that most
people use for general relativity is indeed a 16-dimensional
sub-algebra of the Dirac algebra, so, it does not need to be
addressed specifically.

Particular examples of algebras isomorphic to the Dirac algebra are
those used by Rowlands \cite{Rowlands03, Rowlands07} and the author
himself \cite{Almeida04:4, Almeida06:4}. We have shown how to
convert between those two algebras and the Dirac algebra. A slightly
different case occurs with the algebra used by Carroll
\cite{Carroll04}, because this has an overall dimensionality of 64.
Here we have shown that some of the proposed equations can be set in
an algebra isomorphic to the previous ones and we presented the
means for converting equations between Carroll's algebra and the
remaining ones.

The choice of a particular algebra is irrelevant from the point of
view of the mathematical validity of equations, but it may make a
significant difference to the perception and comprehension of the
physics behind the equations. Quite often, no single choice of an
algebra offers the definitive approach to an equation. Looking at a
particular problem from different angles usually broadens our
perspective over that problem, so, it makes sense to have equivalent
equations written in varied algebras. However, we need to be able to
convert among algebras in order to unify the various approaches.


\bibliographystyle{unsrtbda}
\bibliography{Abrev,aberrations,assistentes}

\end{document}